\documentclass[preprint, 12pt]{elsarticle}
\usepackage[T1]{fontenc}
\usepackage{amssymb}
\usepackage{amsmath}
\journal{Physica D: Nonlinear Phenomena}

\usepackage{xcolor}
\usepackage[unicode]{hyperref}
\usepackage{graphicx}

\def\eq#1{\begin{equation}#1\end{equation}}
\def\eqs#1{\begin{equation}\begin{split}#1\end{split}\end{equation}}
\def\seqs#1{\begin{equation*}\begin{split}#1\end{split}\end{equation*}}
\def\seq#1{\begin{equation*}#1\end{equation*}}

\def\matrix2#1{\left(\begin{array}{cc}#1\end{array}\right)}

\begin{document}

\begin{frontmatter}



\title{On a class of exact solutions of the Ishimori equation}


\author{Rustem N. Garifullin} %% Author name
\author{Ismagil T. Habibullin}
\affiliation{organization={Ufa Institute of Mathematics, Russian Academy 
of Sciences},%Department and Organization
            addressline={\\112 Chernyshevsky Street}, 
            city={Ufa},
            postcode={450008}, 
            country={Russia}}

%% Abstract
\begin{abstract}
In this paper, a class of particular solutions of the Ishimori equation is found. This equation is known as the spatially two-dimensional version of the Heisenberg equation, which has important applications in the theory of ferromagnets. It is shown that the two-dimensional Toda-type lattice found earlier by Ferapontov,  Shabat and Yamilov is a dressing chain for this equation. Using the integrable reductions of the dressing chain, the authors found an essentially new class of  solutions to the Ishimori equation.

\end{abstract}

%% Keywords
\begin{keyword}
3D lattices \sep Toda type lattices \sep generalized symmetries \sep Darboux integrable reductions \sep Davey--Stewartson type coupled system

%% PACS codes here, in the form: 
\PACS 02.30.Ik \sep 02.30.Jr 

%% MSC codes here, in the form: 
\MSC[2020] 35Q51 \sep 35Q56

\end{keyword}

\end{frontmatter}


\section{Introduction}
It is well known that iterations of the B\"acklund transformation for a nonlinear integrable partial differential equation generate nonlinear integrable differential-difference equations called dressing chains \cite{Levi1981}. Dressing chains are actively used to construct particular solutions of nonlinear equations in 1+1 dimension (see, for instance, \cite{ShabatYamilov91}, \cite{VeselovShabat}). The problem of application of this approach to equations in dimension 1+2 came to the agenda many years ago, it was discussed, in particular, in \cite{LeznovShabatYamilov}, \cite{ShabatYamilov97}. However, due to the difficulties with non-local variables, this problem remained open for a long time.
 In a recent article \cite{HabibullinKhakimova} by A.R.Khakimova and one of the authors, significant progress was made in this direction. It was observed that the nonlocality problem can be overcome by imposing a truncation constraint on the dressing chain that preserves integrability. Within the framework of this approach, a model system of the Davey-Stewartson equation type was investigated that is a symmetry of the two-dimensional Volterra chain. In other words, the Volterra lattice is a dressing chain for this model system. Using an explicit solution for a suitably chosen reduction of the dressing chain, an explicit solution for the model system was found depending on two arbitrary functions.

The results of \cite{HabibullinKhakimova} stimulated our interest in coupled systems of partial differential equations associated with higher symmetries of two-dimensional Toda-type lattices. Some of this kind systems are found in \cite{ShabatYamilov97}. In the article cited, the system is presented that is associated with the second-order generalized symmetry of the lattice
\begin{align}\label{fsy}
u_{n,xy}=u_{n,x}u_{n,y}\left(\frac{1}{u_n - u_{n-1}} - \frac{1}{u_{n+1}-u_n}\right),
\end{align}
found earlier by E.V.Ferapontov \cite{Ferapontov97} and independently by A.B.Shabat and R.I.Yamilov \cite{ShabatYamilov97}.
In this paper we have managed to discover the following important fact. It turned out that one of the symmetries of \eqref{fsy}, rewritten as a coupled system, exactly coincides with the well-known Ishimori equation \cite{i86}
\eqs{S_T=-\frac{a}2 S\times (S_{YY}-\varepsilon^2 S_{XX})+\phi_X S_Y+\phi_Y S_X, %S^2=1, S=(S^{1},S^{2},S^{3})\in \mathbb{C}^3,
\\
\phi_{YY}+\varepsilon^2 \phi_{XX}=\varepsilon^2a S(S_X\times S_Y). \label{ishim}}
Here $S=(S^{1},S^{2},S^{3})\in \mathbb{C}^3$ is a point on the {sphere} defined by the equation $\sum_{j=1}^3 (S^j)^2=1.$ { {Since these two equations depend on different sets of variables, the statement requires some explanation. First, we should pass in equation \eqref{ishim} from the variables $S$, $\phi$, $T$, $X$, and $Y$ to the variables $u$, $v$, $h$, $t$, $x$, and $y$ by virtue of the relations \eqref{proj} and \eqref{ind_var} below. Then the flows corresponding to this equation and equation \eqref{ishim} will commute with each other.}}

Recall that \eqref{ishim} is a spatially two-dimensional generalization of the Heisenberg model, which is known for its important applications in the theory of ferromagnets (see, for instance, \cite{Borisov23}, \cite{Kiselev24-1}, \cite{Kiselev24-3}). 

From the above observation it immediately follows that lattice \eqref{fsy} provides a dressing chain for the Ishimori equation rewritten as a coupled system. We use this fact to construct solutions of equation \eqref{ishim}. At first we impose truncation conditions at two points: { $n=0$ and $n=3$} and reduce the dressing chain to a finite-field integrable system. Due to appropriate choice of the cut-off constraint the obtained system is solved in an explicit form \cite{d10}. 
The next and final step in our scheme is to determine the dependence of the found solution of the dressing chain on time $t$ and obtain a solution to \eqref{ishim}.

Let us briefly discuss the content of the article. In \S2 we explain the concept of coupled system associated with lattices and give some examples. In \S3 the Ishimori equation is reduced to a coupled system associated with the lattice \eqref{fsy}. Further, in section 4 general solution for the finite reduction of the lattice is discussed. In \S5 a new explicit solution to the Ishimori equation is found depending on four arbitrary functions. 

It should be noted that the Ishimori equation was previously studied in many papers.  An algorithm for applying the inverse scattering problem method to construct solutions to this equation was discussed (see, for example, \cite{Konopelckenko}, \cite{Imai}, \cite{Habibullin1992}  and references therein). Dynamics of its soliton-type solutions was investigated. A method based on the concept of the Lie point symmetries to this equation is considered in \cite{s13}.

\section{Coupled systems associated with lattice \eqref{fsy}}

\subsection{What is the coupled system associated with a lattice?}

The second-order symmetries of lattice \eqref{fsy} first appeared in  \cite{ShabatYamilov97}. They were derived from the symmetries of the Volterra chain using Miura transformations. It was also noted there that a suitably chosen symmetry transforms into the Heisenberg model when the reduction condition $x=y$ is imposed. However, the relation between the symmetries of lattice \eqref{fsy} and the Ishimori equation had not been noted previously.

The article \cite{HabibullinKhakimovaRCD} presents the symmetries of the lattice \eqref{fsy}, rewritten as coupled systems for the variables $u:=u_n$, $v:=u_{n-1}$. The coupled system in $x$-direction has the form:
\begin{equation}\label{coupledx}
\begin{aligned}
&u_{{x}_2}=u_{{x}{x}}+2u_{{x}}{H},\\
&v_{{x}_2}=-v_{xx}+2v_{{x}}{H}-\frac{2v_{{x}}^2}{u-v}+\frac{ 2v_{{x}}u_{{x}}}{u-v},\\
&D_{{y}}{H}=- D_{{x}} \frac{u_{{y}}}{u-v}
\end{aligned}
\end{equation} 
and the system in the $y$ direction is
\begin{equation}\label{coupledy}
\begin{aligned}
&u_{{y}_2}= -u_{yy}+2u_{y}{Q}+\frac{2u^2_{y}}{u-v}-\frac{2u_{y}v_{y}}{u-v},\\
&v_{{y}_2}=v_{yy}+2v_{y}{Q},\\
&D_{x}Q=D_{y}\left(\frac{v_{x}}{u-v}\right).
\end{aligned}
\end{equation} 
A characteristic feature of the coupled systems above is that all their dynamical variables $u$, $v$, $H$ and $Q$ are taken at the same value of  $n$, therefore, in a coupled system the variable $n$ is an inactive parameter. The concept of coupled system corresponding to the context of the article was first introduced by A.B.Shabat and R.I.Yamilov in their paper \cite{ShabatYamilov97}. Let us illustrate the algorithm for the transition from a lattice symmetry to the associated coupled system using the following system
\begin{equation}\label{shortx1}
\begin{aligned}
&u_{n,x_2}=u_{n,xx}+2u_{n,x}{H}_n, \\
&{H}_n=(T-1)^{-1}D_{x}\log \frac{u_{n,x}}{u_{n+1}-u_n},\\
&D_{y}{H}_{n}=- D_{x} \frac{u_{n,y}}{u_{n}-u_{n-1}},
\end{aligned}
\end{equation}
as an example, here and below $T$ stands for the shift operator acting as $Tf_n=f_{n+1}$. This system is given in \cite{HabibullinKhakimova} (see equation (3.7) with the reduction on p.860) and defines a symmetry of lattice \eqref{fsy}.

Let us show that systems \eqref{coupledx} and \eqref{shortx1} are equavalent.
To this end we rewrite the first and second equations in \eqref{shortx1} in the following form
$$u_{n-1,x_2}=u_{n-1,xx}+2u_{n-1,x}{H}_{n-1},$$
$$H_{n-1}=H_n-D_{x}\log \frac{u_{n-1,x}}{u_{n}-u_{n-1}}.$$ 
Afterwards we exclude function ${H}_{n-1}$ from the first relation by virtue of the second one, and also replace $u_{n-1}$ with $v$ everywhere. As a result, we obtain a coupled system
\eqref{coupledx}. 

Similarly, the symmetry of the lattice  \eqref{fsy} in $y$-direction
\begin{equation}\label{shorty1}
\begin{aligned}
&u_{n,y_2}=-u_{n,yy}+2u_{n,y}{Q}_n+\frac{2u_{n,y}^2}{u_{n}-u_{n-1}}-\frac{2u_{n,y}u_{n-1,y}}{u_{n}-u_{n-1}} , \\
&{Q}_n=-(T-1)^{-1}D_{y}\log \frac{u_{n,y}}{u_{n}-u_{n-1}},\\
&D_{x}{Q}_{n}= D_{y} \frac{u_{n-1,y}}{u_{n}-u_{n-1}}
\end{aligned}
\end{equation}
converts into a coupled system \eqref{coupledy}.

\subsection{A coupled system  symmetric with respect to  $x$ and $y$}

Let us reduce the coupled systems \eqref{coupledx} and \eqref{coupledy} to a more convenient form. We transform them by setting
\begin{align*}
H=K-\frac{u_x}{u-v},\qquad Q=L+\frac{v_y}{u-v}.
\end{align*}
Then the system \eqref{coupledx} takes the form
\begin{equation}\label{coupledx2}
\begin{aligned}
&u_{x_2}=u_{xx}-\frac{2u_x^2}{u-v} +2u_xK,\\
&v_{x_2}=-v_{xx}-\frac{2v_x^2}{u-v} +2v_xK,\\
&K_y=\frac{u_xv_y-u_yv_x}{(u-v)^2}.
\end{aligned}
\end{equation} 
In a similar way one converts \eqref{coupledy} to the form
\begin{equation}\label{coupledy2}
\begin{aligned}
&u_{y_2}=-u_{yy}+\frac{2u_y^2}{u-v} +2u_yL,\\
&v_{y_2}=v_{yy}-\frac{2v_y^2}{u-v} +2v_yL,\\
&L_x=\frac{u_xv_y-u_yv_x}{(u-v)^2}.
\end{aligned}
\end{equation} 

Let us introduce the third system into consideration, taking a linear combination of  two systems above: \eqref{coupledx} and \eqref{coupledy}. The new time $t$ will then be determined by the rule
\begin{align*}
\frac{\partial}{\partial t}=\varepsilon_1\frac{\partial }{\partial x_2}+\frac{\partial}{\partial y_2}, \quad\mbox{where}\quad \varepsilon_1=\pm 1.
\end{align*}
Assuming $h_x=-K$ and $h_y=-L$ we introduce a new non-local variable $h$.  This can be done since the condition $K_y=L_x$ is satisfied. As a result, we arrive at a coupled system of the following form
\begin{equation}\label{coupledfinal}
\begin{aligned}
&u_{t}=\varepsilon_1 u_{xx}-u_{yy}-2\frac{\varepsilon_1 u_x^2-u_y^2}{u-v} -2\varepsilon_1 u_xh_x-2 u_y h_y,\\
&v_{t}=-\varepsilon_1 v_{xx}+v_{yy}- 2\frac{\varepsilon_1 v_x^2-v_y^2}{u-v}-2\varepsilon_1 v_xh_x-2v_yh_y ,\\
&h_{xy}=\frac{u_yv_x-u_xv_y}{(u-v)^2}.
\end{aligned}
\end{equation}

\section{On the relation between the Ishimori equation \eqref{ishim} \\and the lattice \eqref{fsy}}

Let us move from the spherical coordinates of the point $S$ to the coordinates of its stereographic projection:\eq{S^{1}=-\frac{vu-1}{u-v},\quad S^{2}=i\frac{vu+1}{u-v},\quad S^3=\frac{u+v}{u-v}.\label{proj}} 
We introduce the notation $$\phi=-a\varepsilon h, $$
{{ These formulas define the transformation of the sought functions $S^{1},S^{2},S^{3},\phi$ into new functions $u,v,h$. The independent variables are transformed as follows:  }}
% where unknown functions $u,v,\phi$ depend on the variables $t,x,y$ defined as follows: 
\eq{\label{ind_var} t=aT,\quad x=Y-\frac{iX}{\varepsilon},\quad y=-Y-\frac{iX}{\varepsilon}.}
As a result,  equation \eqref{ishim} is rewritten exactly in the form \eqref{coupledfinal} with $\varepsilon_1=-1$:
\eqs{u_t=&-u_{xx}-u_{yy}+2(u_x h_x-u_yh_y)+2\frac{u_x^2+u_y^2}{u-v},\\
v_t=&v_{xx}+v_{yy}+2(v_xh_x-v_yh_y)+2\frac{v_x^2+v_y^2}{u-v},\\
&h_{xy}=\frac{v_xu_y-v_yu_x}{(v-u)^2}.\label{lan_2d}}

%It is easy to see that the systems of equations \eqref{coupledfinal} for $\varepsilon_1=-1$ and \eqref{lan_2d} differ only by insignificant re-notations $\tau=-t$, $h=-\phi$.
This {{observation}} implies the following important fact: the Ishimori equation is a symmetry of chain \eqref{fsy}. Chain \eqref{fsy}, in turn, equipped with the formula for transforming nonlocalities
\begin{equation}\label{Backlund}
\begin{aligned}
&u^*=v\\
&v^*=v-\frac{v_xv_y(u-v)}{v_xv_y +v_{xy}(u-v)},\\
&h^*=h+\ln\frac{v_x}{v_y}
\end{aligned}
\end{equation} 
defines an invertible B\"acklund transformation for the system \eqref{lan_2d}, that in fact performs a shift in $n$, acting according to the rule
$$(v_n, u_n, h_n)\rightarrow(v_{n-1}, u_{n-1}, h_{n-1}).$$
Transormation \eqref{Backlund} can be considered as discrete point symmetry of system \eqref{lan_2d}, however it is a more interesting object since it is invertible by virtue of the lattice \eqref{fsy}. Class of invertible B\"acklund transformations (dressing chains) have earlier been  studied in a series of papers by A.B. Shabat and R.I. Yamilov (see, for instance, \cite{ShabatYamilov91}).

An iteration of mapping \eqref{Backlund}  defines a dressing chain, which can be interpreted as a symmetry of system \eqref{lan_2d}  with discrete ``time''.
In other words, the flows defined by these two systems (with continuous and, respectively, discrete time) commute. Thus, the solutions of the lattice can be used to construct solutions of \eqref{lan_2d}. Usually in practical applications, the stationary part of the symmetry of the equation is taken to construct solutions. Below in the next section we will use other considerations.

\section{Exact particular solutions of the dressing chain}

In order to find a specific solution of the lattice \eqref{fsy} we use  cut-off boundary conditions. Construction of exact solutions of Toda type finite lattices is a very old problem, dating back to classical works of Darboux \cite{d15}. Revival of interest in these problems has occurred in modern literature due to the theory of solitons, see \cite{m79,LSS82}. In relation to the finite field reductions of systems \eqref{fsy}, the problem of constructing solutions is solved in the article \cite{d10}.% see also \cite{d24}.

Namely, we use the following observation. If at some point $n_0$ the sought function takes the same value $u_{n_0}(x,y)=const$ for all $x$ and $y$, then we arrive at two completely unrelated semi-infinite chains located on opposite sides of the point $n=n_0$. Similarly, we can choose two points $n=0$ and $n=m+1$, at which we impose the same boundary conditions of the form $u_{0}(x,y)=c_0$ and $u_{m+1}(x,y)=c_{m+1}$, where $c_0$ and $c_{m+1}$ are arbitrary constants. Next, we will focus on the part of the chain located on the segment $[1,m]$ of the integer line. For $n<0$ and $n>m+1$, we choose the functions $u_n(x,y)$ as identical constants, assuming $u_n(x,y)=c_n$ for any $x$ and $y$. We take the constants $c_n$ so that the inequality
$u_k-u_{k-1}\neq0$ holds for all $k$, that ensures the correctness of the division operation used in the chain.

On the segment $[1,m]$ we do not impose any restrictions on the function $u_{n}(x,y)$. The resulting finite-field reduction of the lattice \eqref{fsy} has the form
\begin{eqnarray}\label{fsyred}
&&u_{1,xy}=u_{1,x}u_{1,y}\left(\frac{1}{u_1-c_0}-\frac{1}{u_{2}-u_{1}}\right),\nonumber\\
&&u_{2,xy}=u_{2,x}u_{2,y}\left(\frac{1}{u_{2}-u_{1}}-\frac{1}{u_{3}-u_{2}}\right), \nonumber \\
&&......................................... \label{rFSY}\\
&&u_{m,xy}=u_{m,x}u_{m,y}\left(\frac{1}{u_{m}-u_{m-1}}-\frac{1}{c_{m+1}-u_{m}}\right). \nonumber 
\end{eqnarray} 
It has a number of advantages, for example, it admits complete sets of characteristic integrals in both directions, i.e. it is a system integrable in the Darboux sense. Recall that complete sets of characteristic integrals reduce a hyperbolic system of PDE to two systems of ODE in $x$ and $y$, that significantly simplifies the problem of constructing solutions.

It should be noted that the above-proposed rule of representing the infinite chain  as three independent subsystems (defined on the intervals $n\in(-\infty,0]$, $n\in[1,m]$ and $n\in[m+1,+\infty)$) is completely consistent with the dynamics in $t$ due to  system \eqref{lan_2d}. We mean that the commutativity condition of the phase flows specified by chain \eqref{fsy} and system \eqref{lan_2d} is not violated. At points $n$ located on the half-lines $(-\infty,0]$ and $[m+1,+\infty)$, the functions $u_{n}(t,x,y)=c_n$ satisfy both systems \eqref{fsy} and \eqref{lan_2d}, i.e. they correspond to a stationary solution.

Below we focus on the simplest non-trivial reduction of the dressing chain
\eqs{v_{xy}=&v_xv_y\left(\frac{1}{v-c_0}-\frac{1}{u-v}\right),\\u_{xy}=&u_xu_y\left(\frac{1}{u-v}-\frac 1{c_3-u}\right)\label{uvxy_1},}
obtained from system \eqref{fsyred} with $m=2$. Here $u:=u_2$ and $v:=u_1$ are the sought functions, and $c_0$ and $c_3$ are constant parameters.
Now we demonstrate derivation procedute of explicit solutions of system \eqref{uvxy_1}, which has been found earlier in \cite{d10}. {This system has characteristic integrals  in $x$-direction
\begin{equation}\label{xintegr}
I_1=\frac{u_xv_x}{(v-c_0)(u-v)(u-c_3)},\quad I_2=\frac{u_{xx}}{u_x}+\frac{2u_x}{c_3-u}+\left(\frac{1}{u-v}+\frac{1}{v-c_0}\right)v_x,
\end{equation}
also integrals in the direction of $y$:
\begin{equation}\label{yintegr}
J_1=\frac{u_yv_y}{(v-c_0)(u-v)(u-c_3)},\quad J_2=\frac{u_{yy}}{u_y}+\frac{2u_y}{c_3-u}+\left(\frac{1}{u-v}+\frac{1}{v-c_0}\right)v_y.
\end{equation}
Let us recall that the integrals are determined from the following relations $D_y I_k=0$, $D_x J_k=0$, $k=1,2$, where the derivatives are calculated by virtue of system \eqref{uvxy_1}.}

{One can see by direct calculations that} system \eqref{lan_2d}  is invariant under the M\"obius transformation. Let us take advantage of this freedom and make the following change of variables
\seq{\tilde u= \frac{c_3(u-c_0)}{u-c_3},\quad \tilde v= \frac{c_3(v-c_0)}{v-c_3},}
($\tilde u,\tilde v$ are new unknowns). This transformation brings the system to a more convenient form
\eqs{v_{xy}=&v_xv_y\left(\frac{1}{v-u}+\frac 1{v}\right),\\u_{xy}=&u_xu_y\ \frac{1}{u-v}\label{uvxy}.}
Here, for simplicity, tildes are omitted. Note that system \eqref{uvxy} is compatible with the Ishimori equation \eqref{lan_2d} as well. Therefore we use  \eqref{uvxy} instead of \eqref{uvxy_1} in our further considerations.

The transformation above significantly simplifies the integrals, which take the form
%Система \eqref{uvxy} является интегрируемой по Дарбу, ее интегралы в $x$ и $y$ направлениях имеют вид:
$$W_1(x)=\frac{u_xv_x}{u-v},\quad W_2(x)=\frac{u_{xx}}{u_x}+\frac{uv_x}{v(u-v)},\quad D_y W_1=D_y W_2=0,$$
$$W_3(y)=\frac{u_yv_y}{u-v},\quad W_4(y)=\frac{u_{yy}}{u_y}+\frac{uv_y}{v(u-v)},\quad D_x W_3=D_x W_4=0.$$
Eliminating the derivatives $v_x$ and $v_y$ from these expressions, we obtain a system containing a single unknown $u$ and four arbitrary functions $W_1(x),$ $W_2(x),$ $W_3(y),$ $W_4(y)$:
\eqs{u_{xx}-W_2(x)u_x+W_1(x)u=0,\\ u_{yy}-W_3(y)u_x+W_4(y)u=0.} Each of which is a second-order linear equation. The coefficients of these equations are arbitrary functions, since they are characteristic integrals of the system. Now it is evident that common solution of the equations \eqref{uvxy} can be represented in the following form: $$u=X_1(x)Y_1(y)+X_2(x)Y_2(y), $$ where $X_1,X_2$ are solutions to the first equation, and correspondingly  $Y_1,Y_2$ are solutions to the second equation. Let us rewrite the latter as
$$u=e^{\ln X_1(x)+\ln Y_1(y)}+e^{\ln X_2(x)+\ln Y_2(y)}$$
and demonstrate that function $u$ is represented in the following compact form: \eq{\label{su}u=w+z,} where functions $w$ and $z$ are arbitrary solutions of the ``wave" equation \eq{(\ln w)_{xy}=0,\quad (\ln z)_{xy}=0.\label{volna}} Now we find $v$ from the second equation in \eqref{uvxy}, and then replace $u$ due to representation \eqref{su}. As a result we get
\eq{\label{sv}v=\frac{(wz_y-zw_y)(wz_x-zw_x)}{zw_xw_y +wz_xz_y }.} As a matter of fact relations \eqref{su} and \eqref{sv} are differential substitutions that map arbitrary solutions of the ``wave" equations \eqref{volna} into solutions of system \eqref{uvxy}.

\section{Solutions to the Ishimiri equation}
Now we proceed to find a solution to the Ishimori equation \eqref{lan_2d} with special forms \eqref{su} and \eqref{sv} of the functions $u,v$.

At  first we search for function $h$. It is noteworthy that in terms of $w,z$ the non-local function $h$ can be defined explicitly. Let us substitute the expressions for $u,v$ into the right-hand side of the equation for $h$ and obtain:
\seq{h_{xy}=\frac{(zw_x-wz_x)(z_{yy}w_y-w_{yy}z_y)}{wz(w_y+z_y)^2}-\frac{(zw_y-wz_y)(z_{xx}w_x-w_{xx}z_x)}{wz(w_x+z_x)^2}} Obviously, it contains functions $w,z$, as well as their derivatives up to the second order.
Therefore, one can look for  function $h$  in the following form: $h=H(w,z,w_x,z_x,w_y,z_y)$. Here we use the concept of dynamical variables of nonlinear PDEs (see, for instance, \cite{LeznovShabatYamilov}). Within the symmetry approach dynamical variables are considered as independent. So after substitution of this representation into the equation for $h$ we get a relation 
containing  dynamical variables $w_{xx},w_{yy},z_{xx},z_{yy}$.
By comparing coefficients in front of these variables we sequentially determine $h$:\eq{\label{sphi} h=\ln\frac{w_x+z_x}{w_y+z_y}.}

There is freedom in defining $h$. It is defined up to two arbitrary functions, one of which depends on $x$ and another on $y$. We choose them equal to zero.

Let us derive differential equations describing the dependence of the functions $w$ and $z$ on t. To this end we substitute expressions \eqref{su}, \eqref{sv} and \eqref{sphi} into equations \eqref{lan_2d}.

After substitution, the first equation is reduced to the form:
\eq{w_t+z_t=w_{xx}+w_{yy}+z_{xx}+z_{yy}.}
Let us express the derivative $w_t$ from this equation $$w_t=w_{xx}+w_{yy}+z_{xx}+z_{yy}-z_t.$$
Then we substitute it into the second equation in \eqref{lan_2d} and obtain a  relationship containing the variables $w, z, z_t,$ $z_{tx}, z_{ty}, w_x, w_y, z_x, z_y, z_{xx}, z_{yy}, z_{xxx}, z_{yyy}.$
Now it is clear that function $z$ satisfies some equation of the form \eq{z_t=H(w,z,z_x,z_y,z_{xx},z_{yy}).} 
Let us explain schematically our further reasoning. We substitute this representation into the relation obtained above and arrive at a certain equation for finding the unknown function $H$. It is important that this equation is overdetermined since it contains some additional independent variables $w_{x},w_{y},z_{xxx},z_{yyy}$. This circumstance allows us to find the explicit form of the desired function $H=z_{xx}+z_{yy}$. Hence we have
\eq{\label{zt}z_t=z_{xx}+z_{yy},} then $w_t$ takes the form: \eq{w_t=w_{xx}+w_{yy}.\label{wt}} 
Thus, each of the functions $w,z$ satisfies the ``wave" equation \eqref{volna} and the heat equation \eqref{zt},\eqref{wt}. 
Due to \eqref{volna} the functions $w,z$ have the following form: \eq{\label{prqs}{w= p(x,t) r(y,t),\quad z= q(x,t) s(y,t).}}
To specify the dependence of these functions on time $t$, we substitute \eqref{prqs} into the equations \eqref{zt}, \eqref{wt} and separate the variables (here we focus only on \eqref{zt} since \eqref{wt} is studied in a similar way)
\eq{\frac{q_t}{q}-\frac{q_{xx}}{q}=-\frac{s_t}{s}+\frac{s_{yy}}{s}=:\omega(t),} where $\omega$ depends only on $t$. Thus we arrive at equations
\eq{q_t-q_{xx}=\omega(t)q,\quad s_t-s_{yy}=-\omega(t)s.}
We remove the terms on the right by linear transformations $q\rightarrow \exp(\int \omega(t)dt)\bar q,$ $s\rightarrow \exp(-\int \omega(t)dt)\bar s.$ Obviously  function $z$ is invariant under this transformations.

 As a result of these considerations, we obtain one-dimensional heat equations \eq{\label{teplo}p_t=p_{xx},\quad q_t=q_{xx},\quad r_t=r_{yy},\quad s_t=s_{yy}} for the sought functions $p,q,r,s.$ 
Now we are ready to write down explicit expressions for the solution $u,v,h$ of the system \eqref{lan_2d} in terms of the solutions of the heat equation:
\eq{\label{soluv}u=pr+qs,\quad v=\frac{(rs_y-sr_y)(pq_x-qp_x)}{q_xs_y+p_xr_y},\quad h=\ln\frac{rp_x+sq_x}{qs_y+pr_y}.} 

Passing from flat coordinates $(u, v)$ to curvilinear coordinates $(S^{1},S^{2},S^{3})$ we obtain an explicit solution to the Ishimori equation \eqref{ishim}:
\eqs{\label{sols}S^{1}=&pr+qs+\frac{((pr+qs)^2-1)(q_xs_y+p_xr_y)}{(sq_x+rp_x)(qs_y+pr_y)},\\ S^{2}=&-i(pr+qs)+i\frac{((pr+qs)^2+1)(q_xs_y+p_xr_y)}{(sq_x+rp_x)(qs_y+pr_y)},\\ S^3=&1+\frac{2(rs_y-sr_y)(pq_x-qp_x)}{(sq_x+rp_x)(qs_y+pr_y)},\\ \phi=&-a\varepsilon\ln\frac{rp_x+sq_x}{qs_y+pr_y}, }
that contains four arbitrary function of one variable, as general solution of any of equation \eqref{teplo} contains one arbitrary function. It can be checked by a direct computation that functions \eqref{sols} satisfy Ishimori equation \eqref{ishim} where functions $p,q,r,s$ solve \eqref{teplo} and $T,X,Y$ are defined by \eqref{ind_var}.

\subsection{Numerical simulation}

We use initial conditions for eqs.\eqref{teplo} of the form 
\seqs{p|_{t=0}=1+\arctan e^x,\quad q|_{t=0}=\frac52+3\arctan e^x,\\ r|_{t=0}=1+2\arctan e^y,\quad s|_{t=0}=2+\arctan e^y.}
In this case one can obtain numerical solutions of \eqref{teplo} which give a global solution for \eqref{ishim}. In the next plot we show the first component of the vector $S^{1}$.
\begin{figure}[h]%% placement specifier
\centering%% For centre alignment of image.
\includegraphics[width=.45\linewidth]{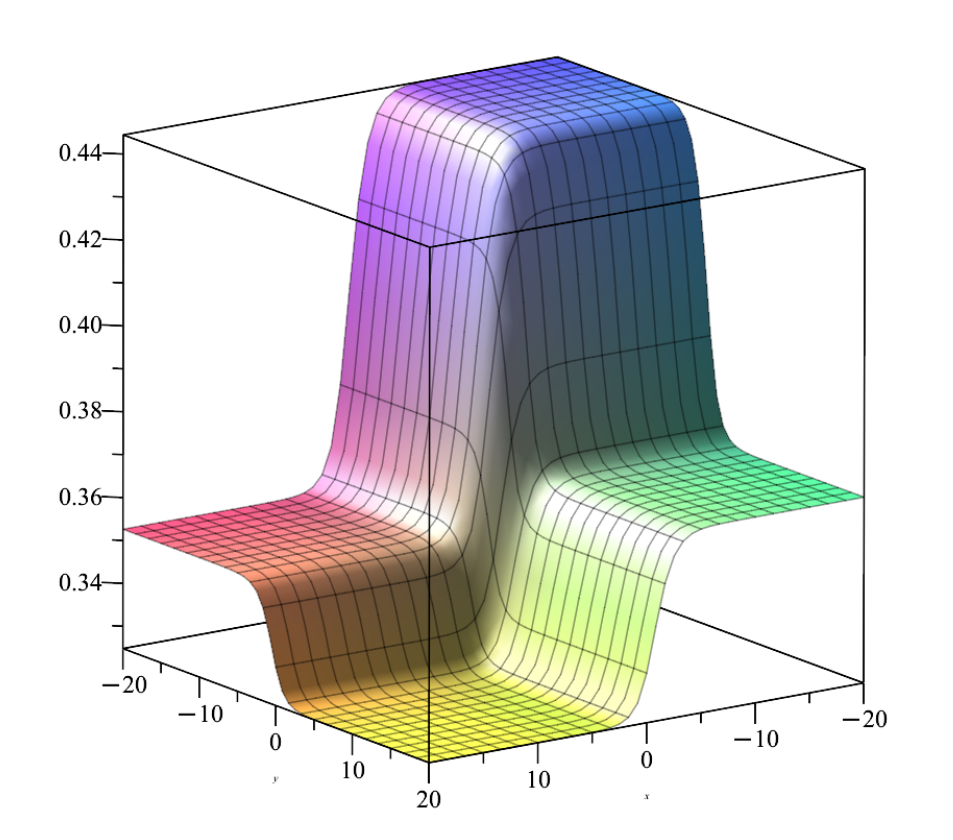} \includegraphics[width=.45\linewidth]{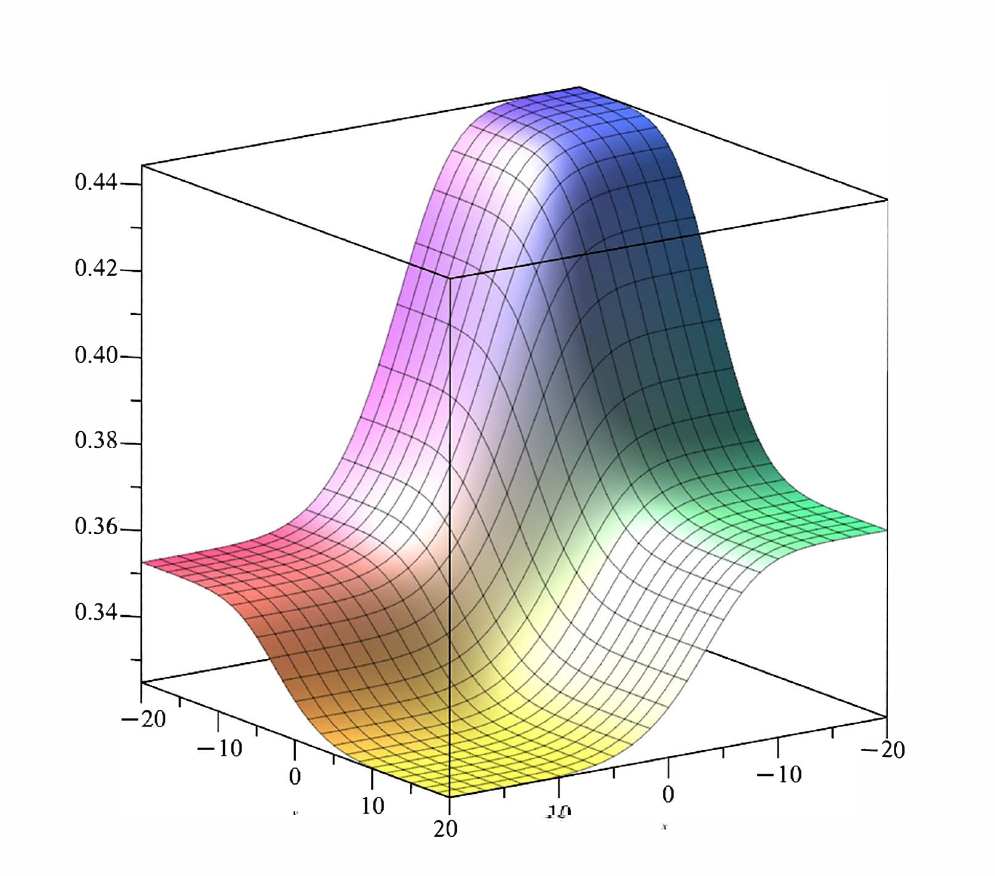}
\caption{$S^{1}$ for $t=0$ and for $t=10$}\label{fig0} 
\end{figure}

\section*{Conclusion}

The class of nonlinear Schr\"odinger type equations and their spatially two-dimensional analogues is of undoubted interest from the point of view of applications in physics (see, for example, recent papers devoted to hydrodynamics \cite{GrinevichSantini} and the dynamics of ferromagnets \cite{Borisov23, Kiselev24-1, Kiselev24-3}). It is well known that multidimensional integrable models are significantly more complex objects compared to equations of dimension 1+1 and therefore they require the use of fundamentally new ideas and approaches (see, for instance, \cite{Imai, Pogrebkov, Willox, BogdanovKonopelchenko, Ferapontov, Taimanov21} and the references therein).
In this paper we constructed a new class of solutions to the Ishimori equation using the dressing chain method. It is worth noting that the solutions found are not related neither to the discrete spectrum of Lax operators nor to classical Lie symmetries.

\section*{Acknowledgments}
The work of R.N. Garifullin is made in the framework of executing the Developing Program of
Scientific and Educational mathematical center of Privolzhsky Federal District (agreement no. 075-02-2025-1637).

%Example citation, See \cite{lamport94}.


\begin{thebibliography}{00}

%% For numbered reference style

%\bibitem{lamport94}
%  \textit{\LaTeX: a document preparation system},
%  Addison Wesley, Massachusetts,
 % 2nd edition,
 % 1994.
\bibitem{Levi1981}  D. Levi, \textit{Nonlinear differential-difference equations as B\"acklund transformation}// J. Phys. A.: Math. Gen, 1981. Vol.14, P.10S3--l098.
\bibitem{ShabatYamilov91} A. B. Shabat, R. I. Yamilov, \textit{Symmetries of nonlinear lattices}, Leningrad Math. J., 2:2 (1991), 377–400
\bibitem{VeselovShabat} A. P. Veselov, A. B. Shabat, \textit{Dressing Chains and Spectral Theory of the Schr\"odinger Operator}// Funktsional. Anal. i Prilozhen., 27:2 (1993), 1–21; Funct. Anal. Appl., 27:2 (1993), 81–96.
\bibitem{LeznovShabatYamilov} A.N. Leznov, A.B. Shabat, R.I. Yamilov. \textit{ Canonical transformations generated by shifts in nonlinear lattices}~// Phys. Lett., A \textbf{174}:5--6, 397--402 (1993).
\bibitem{ShabatYamilov97} A.B. Shabat, R.I. Yamilov. \textit{ To a transformation theory of two--dimensional integrable systems}~// Phys. Lett., A, \textbf{227}:1--2, 15--23 (1997). 
\bibitem{HabibullinKhakimova} I. T. Habibullin, A. R. Khakimova, \textit{Construction of exact solutions of nonlinear PDE via dressing chain in 3D}// Ufa Mathematical Journal, 16:4 (2024), 124–135

\bibitem{Ferapontov97} E.V.Ferapontov, \textit{Laplace transformations of hydrodynamic-type systems in Riemann invariants}// Theoret. and Math. Phys., 110:1 (1997), 68–77.
\bibitem{i86}Yu. Ishimori , \textit{Multi-vortex solutions of a two-dimensional nonlinear wave equation}//Progr.Theor. Phys., 72:1 (1986), 33–37
\bibitem{Borisov23} A. B. Borisov, \textit{Integration of the two-dimensional Heisenberg model by methods of differential geometry}// Theoret. and Math. Phys., 216:2 (2023), 1168–1179  
\bibitem{Kiselev24-1} V. V. Kiselev, \textit{Solitons in a semi-infinite ferromagnet with anisotropy of the easy axis type}// Theoret. and Math. Phys., 219:1 (2024), 576–597
\bibitem{Kiselev24-3} V.V. Kiselev, \textit{Nonlinear dynamics of a two-axis ferromagnet on the semiaxis}// Theoretical and Mathematical Physics, 2024, Volume 220, Issue 3, Pages 1440--1470.

\bibitem{Konopelckenko} B.G. Konopelchenko,  B.T. Matkarimov, \textit{ Inverse spectral transform for the Ishimori equation: I. Initial value problem} //J. Math. Phys. 1990. 31 (11), 2737-2746.

\bibitem{Imai} Kenji Imai, \textit{Dromion and Lump Solutions of the Ishimori-I Equation}//
Progress of Theoretical Physics, Volume 98, Issue 5, November 1997, Pages 1013–1023, https://doi.org/10.1143/PTP.98.1013
P 2737-2746.
\bibitem{Habibullin1992} I. T. Habibullin, \textit{Boundary-value problems on the half-plane for the Ishimori equation that are compatible with the inverse scattering method}// Theoret. and Math. Phys., 91:3 (1992), 581–590. 
\bibitem{s13}Song Xu-Xia, \textit{Lie Symmetries of Ishimori Equation}// Commun. Theor. Phys. 59 (2013), 253--256.
\bibitem{HabibullinKhakimovaRCD} I. T. Habibullin, A. R. Khakimova, \textit{Higher symmetries of the lattices in 3D}// Regular and Chaotic Dynamics, 29:6 (2024).
\bibitem{d15}Darboux G. \textit{Leçons sur la th\'eorie g\'en\'erale des surfaces et les
applications g\'eom\'etriques du calcul infinit\'esimal}. Vol. 2. Paris:
Gautier-Villars, 1915.
\bibitem{m79}A.V. Mikhailov, \textit{Integrability of a Two-Dimensional Generalization of the Toda Chain}// Jetp Lett., 30:7 (1979), 414--418.
\bibitem{LSS82}A.N. Leznov, V.G. Smirnov and A.B. Shabat,  \textit{The group of internal symmetries and the conditions of integrability of two-dimensional dynamical systems}// Theor. Math. Phys. 51, 322–330 (1982).
\bibitem{d10}D.K. Demskoi,  Integrals of open two-dimensional lattices. Theor Math Phys 163, 46--–471 (2010).


%\bibitem{HabibullinPoptsova18} M.N. Poptsova, I.T. Habibullin, \textit{Algebraic properties of quasilinear two--dimensional lattices connected with integrability}// Ufa Math. J., 10:3 (2018), 86--105 
%\bibitem{H-Sakieva} I. T. Habibullin, A. U. Sakieva, \textit{On integrable reductions reductions of two-dimensional Toda-type lattices}// Partial Differential Equations In Applied Mathematics, 11 (2024), 100854 , 9 pp. 

\bibitem{GrinevichSantini} P.\,G. Grinevich,  and P.\,M. Santini, \textit{The finite-gap method and the periodic Cauchy problem for (2+1)-dimensional anomalous waves for the focusing Davey--Stewartson 2 equation}// {Russian Math. Surveys}, 2022, vol.\,77, no.\,6, pp.\,1029--1059.
\bibitem{Pogrebkov} A.\,K. Pogrebkov,  \textit{Negative Times of the Davey--Stewartson Integrable Hierarchy}// {SIGMA}, 2021, vol.\,17, Paper No.\,091, 12\,pp.
\bibitem{Willox} C.R. Gilson, J.J.C. Nimmo, R. Willox, \textit{A (2+ 1)-dimensional generalization of the AKNS shallow water wave equation}// Physics Letters A 180 (4-5), 337-345
%\bibitem{Adler24} V.\,E. Adler, 3D--consistency of negative flows, \textsf{arxiv:nlin/2407.09813v1} (13 Jul 2024).

\bibitem{BogdanovKonopelchenko} L.\,V. Bogdanov,  B.\,G. Konopelchenko,  and A. Moro,  \textit{Symmetry constraints for real dispersionless Veselov--Novikov equation}// {J. Math. Sci.}, 2006, vol.\,136, no.\,6, pp.\,4411--4418.

\bibitem{Ferapontov} E.\,V. Ferapontov,  K.\,R. Khusnutdinova,  and M.\,V. Pavlov,  \textit{Classification of integrable (2+1)-dimensional quasilinear hierarchies }// {Theoret. and Math. Phys.}, 2005, vol.\,144, no.\,1, pp.\,907--915. 

\bibitem{Taimanov21}  I.\,A. Taimanov,  \textit{The Moutard transformation for the Davey--Stewartson II equation and its geometrical meaning}// {Math. Notes},  2021, vol.\,110, no.\,5, pp.\,754--766.


\end{thebibliography}



\end{document}